\documentclass{article}

\usepackage{arxiv}

\usepackage[utf8]{inputenc} 
\usepackage[T1]{fontenc}    
\usepackage[bookmarks=false]{hyperref}       
\usepackage{url}            
\usepackage{booktabs}       
\usepackage{amsfonts}       
\usepackage{nicefrac}       
\usepackage{microtype}      
\usepackage{lipsum}		
\usepackage{graphicx}
\usepackage{natbib}
\usepackage{doi}

\usepackage{tcolorbox}

\usepackage{xspace}
\newcommand{\al}{\textit{et al.}\xspace}

\title{Towards Logging Noisiness Theory:\\
Quality aspects to characterize\\unwanted log entries}

\date{June 6, 2021}	

\author{
Eduardo Mendes \\
SmArtSE Research Team \\
Université du Québec à Chicoutimi \\
555, boulevard de l’Université, G7H 2B1 \\
Chicoutimi, QC, Canada \\
\texttt{eduardo.mendes-de-oliveira1@uqac.ca} \\
\And

Fabio ~Petrillo \\
SmArtSE Research Team \\
Université du Québec à Chicoutimi \\
555, boulevard de l’Université, G7H 2B1 \\
Chicoutimi, QC, Canada \\
\texttt{fabio@petrillo.com} \\
}

\renewcommand{\headeright}{A preprint: June 6, 2021}
\renewcommand{\shorttitle}{Towards Logging Noisiness}

\hypersetup{
pdftitle={Towards Logging Noisiness Theory:
Quality aspects to characterize unwanted log entries},
pdfsubject={cs.SE},
pdfauthor={Eduardo ~Mendes, Fabio ~Petrillo},
pdfkeywords={Logging Noisiness, Logging, Noisy, Quality Aspects, Software Engineering},
}

\begin{document}
\maketitle

\begin{abstract}
    \textbf{Context:} Logging tasks track the system's functioning by keeping records of evidence 
    that have been analyzed by monitoring and observability activities. 
    For these activities to be effective, it is necessary to consider the quality of the consumed information. 
    \textbf{Problem:} However, the presence of noise ― unwanted information, compromises the log files' quality. The noisiness of a log file can be affected among other things by: (i) the wrong severity log choices, (ii) the production of duplicate entries, (iii) the incompleteness of the information, (iv) the inappropriate format of the entries, (v) the amount of information generated.
    \textbf{Objective:} This work aims to broadly define the concept of noise in the context of logging, proposing the initial steps of Logging Noisiness, a theory on quality aspects to characterize unwanted log entries.
\end{abstract}


\section{Introduction}

Logging tasks record system states and events at various significant points during a software system's execution \citep{du2017deeplog}. The generated information will be consumed in debugging \citep{xu2009detecting}, monitoring \citep{boulon2008chukwa, yuan2012improving}, and anomaly detection \citep{tan2008salsa} activities in systems under development or production, whether by systems or humans. Software engineers use them to monitor, for example, whether the system is working or not, to analyze whether it is on the verge of failing, to identify behavioral anomalies, and to understand particularities during its operation through these data.

However, some situations can bias the outcomes of log consuming processes by bringing \textbf{unwanted information} to their inputs, such as the wrong severity log choices, duplicate entries, the incompleteness of the information, and the inappropriate format of the entries. In data processing activities, data quality assessment impacts the performance, time, and cost of analytical processes and the usefulness of their results \citep{borovina2017visualization}.
Furthermore, large-scale software systems generate an overwhelming amount of log data daily \citep{yao2021improving, davis2016effective}. As the system's scale and complexity increase, the amount of generated data also increases. Consequently, unwanted information can accompany this growth \citep{rao2011identifying}, establishing a large volume of noise, posing major challenges to processes as fault diagnosis and root cause determination \citep{zong2019nowhere}.

The ideia of noisy logs has been discussed in some studies. 
Rao \al \citep{rao2011identifying} discuss ambiguous noisy event error logs generated by noisy faults, as random memory errors, disk errors, network glitter, and configuration errors.
Folino \al \citep{folino2009discovering} propose an algorithm to deal with the presence of noise in a process-oriented system's log data.
Datamaran \citep{gao2018navigating} is a tool that extracts structure from log datasets applicable to the noisy log files. Li \al \citep{li2013challenges} identify that distinguishing useful information from noisy logging in error diagnosis during large-scale distributed system deployments is challenging. 

The nuisance of noise is also present on forums such as Stack Overflow \footnote{\href{https://stackoverflow.com}{https://stackoverflow.com} }, and we highlight excerpts from its discussions about it below.

\begin{quote}
    \textit{"My goal is to \textbf{remove these noisy logs} from the console logs in Android Studio..."} \citep{StackAndroidMonitor}
\end{quote}

\begin{quote}
     \textit{"I didn't want my users to have to add a post\_install in their Podfile \textbf{to silence the noisy logging}, ..."} \citep{StackCocoapods}
\end{quote}

\begin{quote}
     \textit{"You could also consider increasing your log level to Debug. \textbf{Most of the noisy logging} MediaPlayer spits out is logged as verbose."} \citep{StackMediaPlayer}
\end{quote}

\begin{quote}
    \textit{"This is causing \textbf{very noisy log files} which makes it a bit of a pain to dig through and is making them unnecessarily large."}  \citep{StackLogRails}
\end{quote}

\begin{quote}
    \textit{ "The problem we are facing is that \textbf{the logs have devolved into noise}, meaning that they fail to provide quick information about failures and successes."} \citep{StackNoisyLoggingProblem}
\end{quote}

The messages mentioned above bring clues to possible causes of noise: from the misunderstanding of logging practices and tools \citep{StackAndroidMonitor, StackCocoapods}, inadequate selection of severity \citep{StackMediaPlayer}, as well as the consequences arising from the high noise of noise, whether they are unnecessarily large files \citep{StackLogRails}, and time overhead to obtain the necessary information about failures and successes \citep{StackNoisyLoggingProblem}.

The objective of this work is to formally and broadly define, describe and develop the concepts about noise in the context of logging. Although this concept is present in forums and the literature, there is a gap in its formal definition as far as we know. In addition, we realize that the concept is applied differently according to the objective pursued by each work.

In order to fill this gap, we seek, in the theory of signals and systems, ways to define better what noise is in various log contexts, the characteristics that characterize it, as well as the consequences and impacts of its presence. Besides, we investigate the state of the art of Software Engineering papers that relate the concept of noise to logging practices.

Our initial findings lead us to propose the first ideas of Logging Noisiness, an approach/theory to characterize log files based on the presence of unwanted or even unnecessary entries.

This paper is organized as follows. Section 2 defines the concepts of noise in Signal and Systems Theory. Section 3 shows a mapping study design to understand how the literature in Software Engineering has presented the concept of noise in the context of logging, highlighting initial observations of the works. Section 4 details Logging Noisiness. Section 5 synthesizes the future research efforts.

\section{Noise in Signal and Systems Theory}

\textit{Noise} can be defined as \textit{"an unwanted signal that is unintentionally added to a desired signal thereby disturbing the latter"} \citep{van2005introduction}. In other words, any interfering signal tends to obscure and mask the desired signals, known as noise. They can be classified as artificial or naturally occurring, and the sources that give rise to unwanted signals varied \citep{lathi1965signals}.

The quality of the received signal is evaluated through the relative sizes of the desired signal and the unwanted signal - noise. In this case, the ratio between the strength of the message signal and the signal to noise (signal-to-noise ratio) is a good indicator of the quality of the received signal \citep{lathi1965signals}.

\section{Log Noise in Software Engineering Literature}

\subsection{Research questions}

\begin{itemize}
    \item \textit{\textbf{RQ1}. What is the definition of noise in the logging context?}
    \item\textit{\textbf{RQ2.} What are the characteristics that make log entries unwanted?}
    \item \textit{\textbf{RQ3.} What consequences are associated with the presence of noise in a log file?}
    \item \textit{\textbf{RQ4.} What software engineering practices and techniques can provide low-noise logging information?}
    \item \textit{\textbf{RQ5.} What is the signal-to-noise ratio in log files?}
\end{itemize}

\subsection{Search strategy and study selection}
We adopted automated search as the search strategy. According to \citep{kitchenham2007guidelines}, automated search is the most common utilized search strategy to identify relevant studies for a Systematic Mapping or a Systematic Literature Review (SLR). In order to perform an automated search, the first step is the creation of a search query \citep{kitchenham2007guidelines}. Our search query is:

\begin{verbatim}
                   ("noisiness log*" OR "noisy log*" OR "noised log*")
\end{verbatim}

We executed our search query on Scopus digital library considering three metadata fields: title, abstract and keywords.

The inclusion criteria are:
\begin{itemize}
    \item \textbf{IC1:} The paper must be a conference paper or article;
    \item \textbf{IC2:} The study must address logging practices and the concept of noise.
\end{itemize}

The exclusion criteria are:
\begin{itemize}
    \item \textbf{EC1:} The study published as an abstract;
    \item \textbf{EC2:} The study is not written in English;
    \item \textbf{EC3:} The study is a keynote, tutorial, or challenge;
    \item \textbf{EC4:} The study does not present a link between logging practices and noise presence.
\end{itemize}

\subsection{Initial observations}

Some works in the literature report challenges arising from the presence and the importance of studying this subject. Li et al.'s study \citep{li2020qualitative} findings show a relationship between noise and processing extensive log data: excessive logging can generate much noise, making it difficult to find what is actually important for failure diagnosis. Yang \al \citep{yang2021interview} conduct a interview study with 25 developers at a company to understand how they use logs. When asking about quality challenges in log analysis, the noise was the most mentioned challenge, 72\% of respondents remembered.

Literature has also been presenting ideas about what noise is. Folino \al \citep{folino2009discovering} and Cheng \al \citep{cheng2015process} discuss noise in the context of process mining. Second Folino \al \citep{folino2009discovering}, "noise in the log refers to situations where the log incomplete, contains errors or reflects exceptional behaviors." Cheng \al \citep{cheng2015process} described a noisy log as duplicated, incomplete, inconsistent, or that reflects incorrect behavior, also presenting possible causes of its origin.
According to Gao \al \citep{gao2018navigating} noise blocks from log datasets have no structure and are not relevant to the tackled problem.
Rao \al \citep{rao2011identifying} use time series as variables and a similarity calculation function to identify noisy in error logs.

\section{Defining Logging Noisiness}

We raised some hypotheses after crossing the background of signals and systems theory and the initial mapping of the literature. These hypotheses will guide the continuation of our investigation.

\begin{tcolorbox}[colframe=gray!25, coltitle=black, arc=0mm, title=\textbf{Hypothesis 0}]
    The concept of noise's framework from Signal and Systems theory allows us to characterize unwanted log entries, improving quality aspects (severity, readiness, redundancy, parseability) of logging artifacts
\end{tcolorbox}

From hypothesis 0, we derive the first definition:

\begin{tcolorbox}[colframe=gray!45, coltitle=black, arc=0mm, title=\textbf{Definition 1 | Noise}]
    In the logging context, \textit{\textbf{noise}} can be defined as any unwanted data that interferes with the information in log statements.
\end{tcolorbox}

Reflecting on this definition and our investigation, we are not only concerned with noise in isolation. We want to understand the impact of different types of noise to qualify log files in their entirety. We wonder how much noise, how "loud" a log artifact can be. To help us in this reflection, we take the meaning of adjectives related to noise. The following terms are presented according to the Cambridge Online Dictionary \citep{harley2000cambridge}. 

\textbf{Noisy} is:

\noindent\fbox{%
    \parbox{\textwidth}{%
        \textit{\textbf{"making a lot of noise",} adjective (sound)
        \begin{itemize}
            \item "The noisiness of the building work should be over by next week."
        \end{itemize}
        }
    }%
}

\noindent\fbox{%
    \parbox{\textwidth}{%
        \textit{\textbf{"having an unwanted change in signal, especially of an electronic device",} adjective (signal) 
        \begin{itemize}
            \item "a noisy signal"
        \end{itemize}
        }
    }%
}

\textbf{Noisiness} is:

\noindent\fbox{%
    \parbox{\textwidth}{%
        \textit{"\textbf{The quality of making a lot of noise",} noun 
        \begin{itemize}
            \item "The noisiness of the building work should be over by next week."
        \end{itemize}
        }
    }%
}

Considering that the quality of log files can be affected by several aspects, we understand that unwanted data can emerge in different ways, making a lot of noise. Thus, we came to that two definitions:

\begin{tcolorbox}[colframe=gray!45, coltitle=black, arc=0mm, title=\textbf{Definition 2 | Noisiness}]
Noisiness is as a set of quality aspects (severity, readiness, redundancy, parseability) to characterize unwanted log entries.\end{tcolorbox}

\begin{tcolorbox}[colframe=gray!45, coltitle=black, arc=0mm, title=\textbf{Definition 3 | Logging Noisiness}]
Logging Noisiness (LN) is the level of  unwanted/needless (not useful) logging entries in a specific context of a system.
\end{tcolorbox}

Some of the criteria for noisiness that we've been able to list in the research so far are: severity, results/quantity, readiness, data density, redundancy, parseability and statistical deviation. 

Below we present some of the hypotheses we are working on to expand the concept of noisiness. 
\begin{tcolorbox}[colframe=gray!25, coltitle=black, arc=0mm, title=\textbf{Hypothesis 1}]
    Logs with a lot of debug entries in production are noisiness.
\end{tcolorbox}

\begin{tcolorbox}[colframe=gray!25, coltitle=black, arc=0mm, title=\textbf{Hypothesis 2}]
    Logs that have log entries with appropriately classified log levels has a low noisiness level.
\end{tcolorbox}

\begin{tcolorbox}[colframe=gray!25, coltitle=black, arc=0mm, title=\textbf{Hypothesis 3}]
    The noisiness level of logs indicates its anomaly degree.
\end{tcolorbox}

\begin{tcolorbox}[colframe=gray!25, coltitle=black, arc=0mm, title=\textbf{Hypothesis 4}]
    The noisiness of log files reveals developers' lack of knowledge of good logging practices.
\end{tcolorbox}

\begin{tcolorbox}[colframe=gray!25, coltitle=black, arc=0mm, title=\textbf{Hypothesis 5}]
    The noisiness of log files reveals the need for adjustments in operating settings and activities.
\end{tcolorbox}

\section{Final remarks}
The quality of log files impacts the processes that consume them, and several factors can affect this quality, causing the presence of unwanted data - the presence of noise.

In this work, we present the initial stages of building a theory focusing on understanding the presence of noise in logging and the impact on log files as a whole. We present the problem of noise in the log, indicated by the literature and by software practitioners. In addition, we present the study design for our problem mapping and related work.

The related works indicate the importance of studying the theme, which made us raise some hypotheses. We present our hypotheses as well as the first definitions of Logging Noisiness.

\bibliographystyle{unsrtnat}
\bibliography{references}  

\end{document}